\documentclass[12pt]{article}
\usepackage{amsmath,amsfonts,amssymb,a4,epsfig}

\newcommand{\R}{{\mathbb{R}}}
\newcommand{\C}{{\mathbb{C}}}
\newcommand{\Z}{{\mathbb{Z}}}

\def\ha{\frac{1}{2}}

\def\pa{\partial}
\def\ra{\rightarrow}
\def\preuve{\begin{proof}}

\def\ga{\alpha}

\def\ge{\varepsilon}

\def\gf{\varphi}
\def\gg{\gamma}

\def\gl{\lambda}

\def\gs{\sigma}

\def\san{San V{\~u} Ng{\d o}c}
\newtheorem{defi}{Definition}
\newtheorem{lemm}{Lemma}

\newtheorem{rem}{Remark}

\newtheorem{theo}{Theorem}

\newenvironment{demo}{\noindent {\it Proof.--}
      \begin{quotation}\noindent}{\end{quotation}\hfill$\square $}

\def\aujour{\ifnum\day=1 1\ermini\else\number\day\fi\
\ifcase\month\or janvier\or f\'evrier\or mars\or avril\or mai\or
 juin\or
juillet\or ao{\^u}t\or septembre\or octobre\or novembre\or
 d\'ecembre\fi\
\number\year}

\begin{document}

\title{Bohr-Sommerfeld phases for
avoided crossings}
\author{Yves Colin de Verdi\`ere \footnote{Institut Fourier,
 Unit{\'e} mixte
 de recherche CNRS-UJF 5582,
 BP 74, 38402-Saint Martin d'H\`eres Cedex (France);
yves.colin-de-verdiere@ujf-grenoble.fr}
}


\maketitle



\section*{Introduction}

\subsection{Adiabatic limit in quantum mechanics}
The problem of adiabatic limit in quantum mechanics with avoided 
eigenvalues crossings will be the basic example in our paper.

Let us describe it in  a  precise  way.
We start with a smooth  family of $N\times N$ Hermitian matrices 
$A(t)$  with $a\leq t \leq b$
and consider the following linear system of differential equations:
\begin{equation} \label{equa:adia}
\frac{h}{i}\frac{dX}{dt}=A(t) X 
\end{equation}
Solving this equation gives an unitary map 
${\cal S}_h $ defined by
${\cal S}_h(X(a))=X(b) $. We  call it the scattering matrix of the
problem.
Adiabatic Theorems describe the asymptotic behaviour
of ${\cal S}_h $ when $h\ra 0$.

\begin{enumerate}
\item 
The clasical adiabatic theorem  \cite{B-F,Kato} concerns the
 case where the eigenvalues
of $A(t) $ satisfy, for all $t$:
\begin{equation} \label{equa:nogap}
\gl_1(t) < \gl _2 (t)< \cdots < \gl _N (t) ~.
\end{equation}
In this case, if we start with $X_j(a)$ an eigenvector
of $A(a)$ with eigenvalue $\gl_j (a)$, we have
$X_j(b)={\rm exp}(i\int_a^b \gl _j (s) ds )Y_j (b)+O(h)$
where $Y_j(b)$ is obtained by parallel transporting $X_j(a)$
 along $[a,b]$
in the eigenbundle $L_j(t)=\ker (A(t)-\gl_j(t))$ w.r. to the geometric
 (or Berry)
connection
defined
by
 \[\nabla _{\pa /\pa t}Y(t)={\rm proj}_{L_j(t)}\frac{dY}{dt} \]

\item Another case which is well known \cite{A-E,B-F,Hage,Ro}, is the case where the
eigenvalues of $A(t)$ cross transversally. Then the previous
results remains true with a less good remainder term $O(\sqrt{h})$
but we need to label the eigenbundle by following smoothly
the eigenvalues at  the crossings points. In other words, there 
is some exchange between $L_j$ and $L_{j\pm 1 }$ at each
crossing point.

So the scattering matrix remains diagonal but with some
relabeling of the indices $j=1, \cdots, N$.
\end{enumerate}

We want to consider the bifurcation between the 2 situations.
For that purpose, we allow $A(t)$ to depend smoothly of 
a $d-$dimensional parameter $\mu $ close to $0$ in $\R^d$.
We will assume that $A_0 (t)$ admits eigenvalues
crossings transversally. For $\mu $ small, $A_\mu $ will
have {\it avoided} eigenvalue crossings.
We will derive in this paper a way to get the
asymptotic behaviour of ${\cal S}_{\mu,h }$
w.r. to both small parameter $\mu$ and $h$.

Basically the result uses 2 parts :
\begin{enumerate}
\item 
 The local situation   where a suitable Landau-Zener formula
can be used describing the local $2\times 2$ scattering matrix
(see \cite{Jo1,C-L-P}).
\item The global problem where we have to take into account
interferences given by what we called Bohr-Sommerfeld phases.
In the  case of several crossings (analytic case),
 using Stokes lines, they are many works
where the interferences effects are exponentially small
 \cite{Joye,M-N}.
\end{enumerate}

\subsection{Born-Oppenheimer approximation}

Another important example comes from the Born-Oppenheimer approximation.
We consider, on the real line a Schrödinger equation
with matrix valued potential as follows:
\begin{equation} \label{equa:bop}
\hat{H}=-h^2 \frac{d^2}{dx^2}+ V(x) 
\end{equation}
where $V$ is an Hermitian matrix depending smoothly on $x\in \R $.
We are interested in the eigenvalue equation
$( \hat{H}-E)\vec{u}(x)=0 $ 
with $\vec{u}: \R \ra \C^n $.
If there is no degenerate eigenvalues of $V(x)$ smaller than
$E$, then eigenfunctions can be foud using a WKB-Maslov
Ansatz in a very similar way to the scalar case.
It is the generic situation for a single $V$ \cite{Ha-To,Ro2}). The most precise results 
are shown in \cite{E-W}. However, if $V$
depends on some external parameters $\mu $ close
to $0$ in $\R^d$, it is possible that, in a stable way,
$V_\mu (x)$ admits  degenerate eigenvalues.
The problem we adress in this case is the description of
Bohr-Sommerfeld rules uniformly w.r. to $\mu$.

\subsection{The general setting}

Both examples can be put together as follows:
we consider  a self-adjoint semi-classical system of 
$N$ (pseudo-)dif\-fe\-ren\-tial equations  of order $0$
with $N$ complex valued unknown functions
$\vec{U}=(U_1, \cdots, U_N)$,
\begin{equation} \label{equ:1}
\widehat{H}\vec{U}=0(h^\infty)
\end{equation}
on the real line. Here 
$\hat{H}=(\hat{H}_{i,j})_{1\leq i,j \leq N}$ is a matrix of
semi-classical (pseudo-)differential operators of order $0$ in $1$ variable $x$
 with $(\hat{H}_{i,j})^\star=\hat{H}_{i,j}$.
 Viewing the revious ewmples, it will be important to consider
the case where $\hat{H}=\hat{H}_\mu  $ depends smoothly
  on a germ of  parameters
 $\mu \in (\R^d, 0)$.
In our previous papers \cite{CV1,CV2}, we derived normal
 forms near the 
eigenvalues
crossings which allow to compute a {\it local scattering}
 matrix including the
Landau-Zener amplitude.
The goal of this paper is to compute {\it global objects} 
including {\it interferences effects.}
 The general picture is already provided
by the study of the scalar case \cite{CP3} from which we know that
 we need to 
define ad hoc {\it Bohr-Sommerfeld phases.}

The general terminology   is the same as in \cite{CV1} and
 \cite{CV2},
but in the present paper, our phase space will always be 2
 dimensional:
\[ H_{\rm class}^\mu  :T^\star \R \ra {\rm Herm}(\C^N)~,\]
 the 
(matrix valued) principal
symbol, is the  {\it dispersion matrix},
  and $C_\mu=p_\mu ^{-1} (0)$ with
 $p_\mu = {\rm det} (H_{\rm class}^\mu)$ 
the {\it dispersion relation.}

We first recall the local normal form as derived in our previous
papers \cite{CV1,CV2}  and we solve it. After that, we come to the
 new part which 
consists in  deriving global objects in the spirit
of \cite{CP3}.

We will need another piece of information which we call
{\it Bohr-Sommerfeld phase;}
let us take any simple cycle $c$ (with singular vertices
$z_1,\cdots, z_j, \cdots, z_p$) of the dispersion relation
$C_0$. We will associate to $c$
a   real valued symbol
${\bf S}_h(\mu)\sim 
\sum _{j=0}^\infty S_j(\mu) h^j $
were  the $S_j$'s are formal power series in $\mu$.
$S_0$ is a purely classical object which involves
regularized action integrals.
$S_1 (0)$ is computed using the transport equation which
is smooth along the edges $[z_j, z_{j+1}]$ {\it (Berry phases)}
and singular Maslov indices.
From the Bohr-Sommerfeld phases we recover the global objects 
mod $O(h^\infty )$.

\section{The local normal form}

Let us recall the following result from \cite{CVV,CV1,CV2}
(see also \cite{C-L-P}):
\begin{theo}\label{theo:normal}
Let us assume that 
the function $p_0(x,\xi)={\rm det}(H_{{\rm class},\mu= 0})$
admits at the point $z_0 \in T^\star \R $ a
non degenerated  critical
 point of Morse index $1$ (also called {\rm hyperbolic critical
 point},
because the Hamiltonian vector field of $p_0$ is hyperbolic at the
singular point $z_0$)
and with critical value $p_0(z_0)=0$.

Then, we can find the following objects which depends smoothly
 on $\mu$ close
enough to $0$:
\begin{itemize}
\item A smooth  family of  germs of canonical transformations
$\chi _\mu :(T^\star \R , 0) \ra (T^\star \R,z_0) $
such that 
\[ p_\mu (\chi _\mu (x,\xi)) = e_\mu (x,\xi)(x\xi - \gg_0( \mu))
 \]
with $e_\mu $ an invertible germ    function and $\gg_0 $
a    germ of  $\geq 0$ function of $\mu$ satisfying
$\gg_0(0)=0 $. Moreover, the Taylor expansion of $\gg _0$
is unique.
\item A smooth family of unitary FIO's $U_\mu$ associated to
$\chi _\mu$ and $N\times N$ matrix of  $\Psi$DO 's $A_\mu $
\end{itemize}
so that, we have the following normal form
(called the {\rm Landau-Zener normal form})
 near $z_0$:
\[ A_\mu ^\star U_\mu ^\star \hat{H}_\mu  U_\mu A_\mu =
 \left(  \begin{array}{cc}
\left(  \begin{array}{cc}
D & \ga \\
 \bar{\ga} & x
\end{array} \right)&0 \\
0 & Q \end{array} \right) \]
with $D=\frac{h}{i}\frac{\pa }{\pa x}$,
 $\ga (\mu, h) \sim \Sigma _{j=0}^\infty a_j (\mu) h^j $
 a symbol and $Q$ is elliptic.

Moreover, $a_0$ is a complex valued function of $\mu$
which satisfies 
\[ |a_0|^2 (\mu)=  \gg_0(\mu)~, \]
and we have 
\[ \gg_0(\mu) =-
\frac{p_\mu (z_0)}{\sqrt{|{\rm det}p''_0 (z_0)|}  }
+O(\mu ^3) ~.\]

\end{theo}
\begin{rem}\label{rem:transv} In  \cite{CV2},
 Theorem \ref{theo:normal}
 is proved under the
following  transversality hypothesis:

{\bf $(\star )$  if $W \subset {\rm Herm }(\C^N )$ is the
 submanifold
defined by $\dim \ker H =2$, we assumed there
that $(\mu,z) \ra H_{{\rm class},\mu }(z) $ is transversal
to $W$ at the point $(0,z_0)$.}

 This hypothesis  can be restored
using more parameters, so that Theorem \ref{theo:normal}  is
 also correct.
For simplicity,

 {\bf we will assume that hypothesis
$(\star )$ holds true  in what follows.}

It implies that $\mu \ra a_0(\mu) $ is a submersion
from $(\R^d, 0) $ onto $(\C, 0)$.
We will denote by $Z= a_0^{-1}(0)$.
$Z$ is a smooth germ of codimension $2$ manifold of $(\R^d, 0)$.
Formal expansions w.r. to the parameter $\mu$ mean formal
 expansions
along $Z$.

\end{rem}
\begin{rem}\label{rem:orient}
Contrary to the scalar case, there is no arbitrary choice 
concerning the images of the half axes
$\{ \xi=0, x>0\}$, ... by $\chi$.
The smooth arcs of the dispersion relation are oriented 
in the following way: there is a change of the Morse index of the
quadratic form associated to $H_{\rm class}$ from $m$ to 
$m\pm 1$ while crossing these arcs. The sign of this change is
preserved  by the gauge transform which acts directly on the
 previous
quadratic form by an invertible linear change of variable.
We choose to orient the arcs so that the Morse index is bigger
on the right than on the left of the path.
\begin{figure}[hbtp]
  \begin{center}
    \leavevmode
    \input{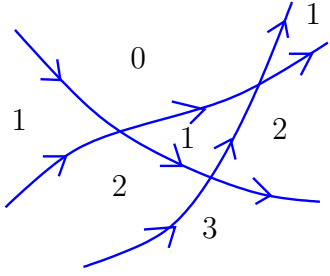}
    \caption{the jumps of the Morse index 
of the dispersion matrix}
    \label{fig:morse}
  \end{center}
\end{figure}

\end{rem}
\begin{rem}\label{rem:unique}
The symbol $\ga $ is not
uniquely defined because
a diagonal unitary gauge transform
preserves the normal form while changing
$\ga$ by some phase shift ${\rm exp}(i\gf (h))$.
Its  modulus $\gg(\mu,h)=|\ga (\mu,h)| ^2 $ is uniquely defined
from the Landau-Zener coefficient given in Equation
(\ref{equ:tau}).

The matrix $A$ is  defined up to   matrices
which will change the phase of $\ga $.
More precisely, if $A_{0}$ is the principal symbol of 
$A$ at the crossing point, the only prescription is that
$A_{0}$ is a  map
from $\C^N$ to $\C^N$ which sends $\C^2 \oplus 0$
into $\ker H_0(z_0) $
and  satisfies
$A_{0}\left((\C \oplus 0 )\oplus 0\right)=E_1$ and
 $A_{0}\left((0 \oplus \C)\oplus 0\right)
=E_2$ where $E_j =\lim_{z\ra z_0, z\in \Lambda _j\setminus z_0}
\ker H_0$
with $\Lambda _1 = \chi _0 (\{\xi=0\})$
and  $\Lambda _2 = \chi _0 (\{x=0\})$.
The choice of $(A_{0})_{|\C^2 \oplus 0}$ will be important
 in the computation
of $S_1(0)$ in section \ref{sec:S1}.

\end{rem}

The previous result is a microlocal result
 and the subject of the present
paper is to get a global result.

In the {\it  adiabatic case} (see Section \ref{sec:adiab}), we get 
\[ \gg_0 (\mu)=\frac{{\rm gap}(\mu)^2}{4( |\gl'_+-\gl'_- |)}+
O(\mu^3) ~\]
where 
${\rm gap}(\mu) $ is the minimal gap of the avoided crossing
and $\gl'_\pm $ are the slopes of the unperturbed eigenvalues
at the crossing point.

\section{The local scattering matrix for the Landau-Zener 
normal form}
\label{sec:local}

The goal of this section is to compute in a very explicit way
the local $2 \times 2$ scattering matrix ${\cal T}$ for the
Landau-Zener
normal form~:
\begin{equation}\label{equ:LZ}
{\rm (LZ)} \left\{  \begin{array}{cc}
Du + \ga v&=0\\
 \bar{\ga}u+ xv &=0 
\end{array} \right. \end{equation}
with $D=\frac{h}{i}\frac{\pa }{\pa x}$.

Let us put $\gg= |\ga |^2 $ and 
let us choose some small $a>0$ and assume
$\gg < a^2$.
Let $C_\ga^{\rm LZ}  =\{ x\xi= \gg \} $ be
the characteristic manifold.
The set $C_\ga ^{\rm LZ} \cap \{  \max( |x|, |\xi|) \geq a \}$
is the union of 4 connected arcs. These arcs are labelled 
$\Lambda _\pm ^{\rm in, out}$ as follows:
\begin{itemize}
\item $\Lambda _ +^{\rm  in} =
\{(x,\xi) \in C_\ga^{\rm LZ} ~|~ x\leq -a  
 \} $
\item $\Lambda _+ ^{\rm out} =\{ (x, \xi) \in C_\ga^{\rm LZ} ~|~
  \xi \leq -a   \} $
\item $\Lambda _- ^{\rm in } =\{ (x,\xi) \in C_\ga^{\rm LZ} ~|~
  \xi \geq a   \} $
\item $\Lambda _ -^{\rm out } =\{(x,\xi)
 \in C_\ga^{\rm LZ}~|~  x \geq
 a 
   \} $.
\end{itemize}

The meaning of the labels is as follows:
\begin{itemize}
\item ``in'' (resp. ``out'')
means that the arc oriented according to
remark \ref{rem:orient}   is incoming (resp. outgoing).
\item ``$+ $'' (resp. ``$-$'')
 means that the vanishing eigenvalue of the dispersion
matrix is the largest (resp. smallest) one.
\end{itemize}

We start defining  4 WKB (exact) solutions of the previous system
associated to the  4 Lagrangian arcs $\Lambda _\pm 
^{\rm in, out}$:

\begin{figure}[hbtp]
  \begin{center}
    \leavevmode
    \input{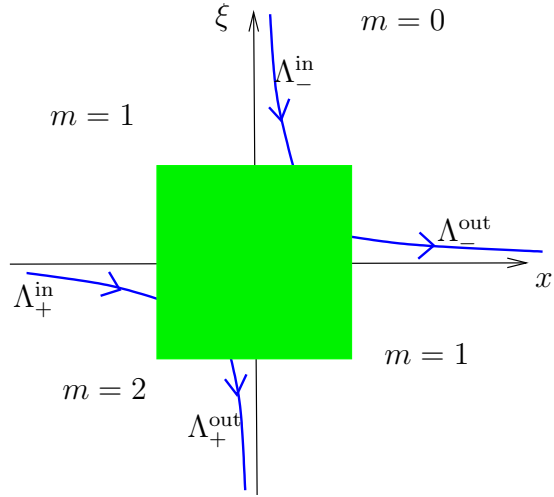}
    \caption{the 4 arcs of the characteristic manifold
and Morse indices}
    \label{fig:normal}
  \end{center}
\end{figure}

\begin{equation} \label{equ:sol_nf} \left\{ \begin{array}{cc}
W_-^{\rm out}~:~&u_-^{\rm out}(x)=x_+^{i\frac{\gg}{h}}   
 ,~v_-^{\rm out}(x)=-\bar{\alpha}x_+^{i\frac{\gg}{h} -1}  \\
W_+^{\rm in}~:~&u_+^{\rm in}(x)= x_-^{i\frac{\gg}{h}}    ,
  ~v_+^{\rm in}(x)=\bar{\alpha}x_-^{i\frac{\gg}{h} -1}   \\
W_-^{\rm in}~:~& \widehat{u_-^{\rm in}}(\xi )=  
- \alpha \xi_+^{-i\frac{\gg}{h}-1}           ,
~\widehat{v_-^{\rm in}(\xi)}=\xi_+^{-i\frac{\gg}{h}} \\
W_+^{\rm out} ~:~& \widehat{u_+^{\rm out}}(\xi)=  \alpha
\xi_-^{-i\frac{\gg}{h}-1}        ,
~\widehat{v_+^{\rm out}}(\xi)=\xi_-^{-i\frac{\gg}{h}} 
\end{array} \right. \end{equation}
where $\widehat{f}(\xi)$
is the $h-$Fourier transform of $f(x)$
defined by
\[  \widehat{f}(\xi)=\frac{1}{\sqrt{2\pi h}}
\int_\R e^{-i\frac{x\xi}{h}}
 f(x) |dx| ~,\]
and $x_\pm =Y(\pm x) |x| $ with $Y$ the Heaviside function.

Computing the Fourier transforms of $u_\pm $ and $v_\pm $,
we get the following compatibility conditions
in order to get microlocal solutions of (\ref{equ:LZ}) near the origin:
\[ \left\{ \begin{array}{cc}
W_-^{\rm out}(x)\leftrightarrow 
&h^{\ha+ i\frac{\gg}{h} }
 \frac{\Gamma (1+i\frac{\gg}{h})}{\sqrt{2\pi}\alpha}
\left( ie^{\pi \frac{\gg}{ 2h}}
W_-^{\rm in}(x)+ ie^{-\pi \frac{\gg}{ 2h}}W_+^{\rm out}(x)
 \right) \\
W_+^{\rm in}(x)\leftrightarrow 
&h^{\ha +i\frac{\gg}{h}}
 \frac{\Gamma (1+i\frac{\gg}{h})}{\sqrt{2\pi}\alpha}
\left(- ie^{-\pi \frac{\gg}{ 2h}}
W_-^{\rm in}(x)-i  e^{\pi \frac{\gg}{ 2h}}W_+^{\rm out}(x)
 \right) 
\end{array} \right. \]

If $ W^{\rm in}:=y_+ W_+^{\rm in} + y_- W_-^{\rm in} $
and $W^{\rm out}:= z_+ W_+^{\rm out} + z_-  W_-^{\rm out}$
are  WKB-solutions of Equation (\ref{equ:LZ})  outside the origin,
we get, for any microlocal solution near the origin, 
\[ \left( \begin{array}{cc} z_- \\ z_+ \end{array}
\right) = {\cal T} \left( \begin{array}{cc} y_+ \\ y_- \end{array}
\right) \]
where $ {\cal T}$ is the {\it unitary matrix} defined by: 
\begin{equation} \label{equ:T}
 {\cal T}=
\frac{1}{A} \left( \begin{array}{cc}
-B & 1\\   B^2-A^2&-B  \end{array} \right)\end{equation}
with
 \begin{equation}\label{equ:AB}
A= i h^{\ha +i\frac{\gg}{h}}
 \frac{\Gamma (1+i\frac{\gg}{h})}{\sqrt{2\pi}\alpha}
  e^{\pi\frac{ \gg}{ 2h}},~
B=ih^{\ha+i\frac{\gg}{h} }
\frac{\Gamma (1+i\frac{\gg}{h})}{\sqrt{2\pi}\alpha }
 e^{-\pi \frac{\gg}{ 2h}} ~.
\end{equation}

The matrix ${\cal T }$ will be called the {\it local
scattering matrix} associated to the singular point (and the choice of a normal
form).
Unitarity of ${\cal T }$ is  checked using 
 the well  known formula 
\[  \Gamma (1+ix)\Gamma(1-ix)= \frac{\pi x}{\sinh \pi x} \]
which implies
$ |A|^2=|B|^2+1,~ A\bar{B}= \bar{A}B$.

The transmission coefficient 
\begin{equation} \label{equ:tau}
 \tau = \left| \frac{B}{A} \right|={\rm exp}(-\pi \frac{\gamma}{h}  )
\end{equation}
gives  the  {\it Landau-Zener formula}.
The previous explicit expression for the scattering matrix 
will allow to define in the next section 
 the Bohr-Sommerfeld phases and to take into account
interferences patterns due to several (avoided) crossings.

\section{Singular Bohr-Sommerfeld phases }

\subsection{Outline}
To each cycle $c$ of the dispersion relation $C_0$, we can
 associate,
 using the
recipe of \cite{CP3}, a singular phase
of the form
\[ {\bf S}_h  (\mu)= S_0 (\mu) +hS_1(\mu) +\cdots \]
where the $S_j$'s are smooth w.r. to $\mu$.

In this section, we will  define precisely these phases. We  show
that they are uniquely defined as formal power series w.r. to
 $(\mu,h )$.
More precisely, each $S_j (\mu)$ is well defined modulo flat
 functions
on $Z$ (see Remark \ref{rem:transv}).
We will  give more precise properties  of $S_0$ in section \ref{sec:S0}
and $S_1$ in section \ref{sec:S1}:
$S_0(\mu)$ is, as a formal power series,
 a purely classical object
derived from the dispersion relation,  while $S_1(0)$ is a
 semi-classical   object  associated to phases given by the 
transport equation which in the adiabatic  case are {\it Berry
 phases}.

\subsection{Bohr-Sommerfeld phases: a definition}

Let us take a simple   oriented cycle $c$
  of the dispersion relation
$C_0$ (boundary of a  bounded
component of $T^\star \R \setminus C_0$).
 Let $z_1,~ z_2, \cdots, z_n $ be the singular
points of $c$ ordered cyclically around $c$.
For each singular point $z_j$, let us build a FIO $U_j$
and a $\Psi DO $ gauge transform $A_j$ (all depending
smoothly on $\mu $) which give the normal form 
of Theorem \ref{theo:normal} with
$\ga _j = \ga _j  (\mu,h)$ a full symbol.

We will define ${\cal H}_\mu(c)= {\rm exp}{(i{\cal S}_h (\mu)/h)}$
as follows:
we will denote by
$W_{\pm,j}^{\rm in, ~out} $
 the images of $W_{\pm}^{\rm in, ~out}  $
 by the 
operators $U_j^\mu A_j^\mu $.
These functions are WKB solutions of equation
(\ref{equ:1}) associated to arcs of $C_\mu $ near $z_j$.
We introduce also WKB solutions $u_j $ of (\ref{equ:1})
along arcs of $C_\mu $ close to $]z_j, z_{j+1}[$.
From those objects we get a global holonomy ${\cal H}_\mu (c)$
of the cycle $c$ defined as follows:
we have (by uniqueness,
modulo  multiplication by a full symbol, of WKB
solutions)  for example near $z_j$:
\[ u_j =x_j W_{- ,j}^{\rm in},~u_{j-1}=y_jW_{+ ,j}^{\rm
in}
  ~. \]
We define
${\cal H}_\mu(c)=\Pi _{j=1}^n y_j x_j^{-1} $.
In other words, ${\cal H}_\mu(c)$ is the holonomy of a sheaf on
$c$ given by the WKB solutions on the smooth part of the cycle
and whose jumps of section are given from the normal forms. In our
previous example $ W_{+ ,j}^{\rm
in}\ra W_{- ,j}^{\rm in}$.

\begin{figure}[hbtp]
  \begin{center}
    \leavevmode
    \input{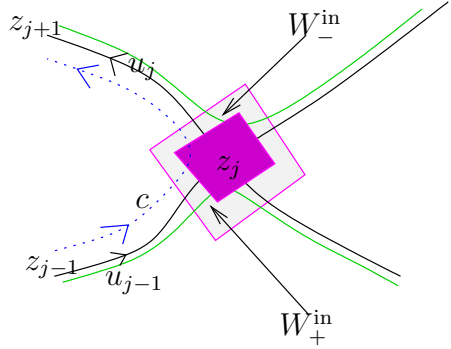}
    \caption{defining ${\cal H}_\mu(c)$}
    \label{fig:holloc}
  \end{center}
\end{figure}

\begin{lemm}
We have 
$|{\cal H}_\mu(c)|=1+O(h^\infty )$.
\end{lemm}
\begin{demo}
Following \cite{CV3} section 11.2.1. and  Figure
\ref{fig:scatt_cycle},
we associate to the cycle $c$ an unitary scattering
 matrix ${\cal S}_c$
which is computable from
the local  unitary scattering matrices associated to the singular
points
and the holonomy ${\cal H}_\mu(c)$.
If this holonomy does not satisfy $|{\cal H}_\mu(c)|=
1+O(h^\infty
)$,
  the global scattering matrix would not be
unitary: the previous matrix is the product of (unitary)
local scattering matrices and a diagonal matrix
whose unique nonzero entry is  ${\cal H}_\mu(c)$.

\begin{figure}[hbtp]
  \begin{center}
    \leavevmode
    \input{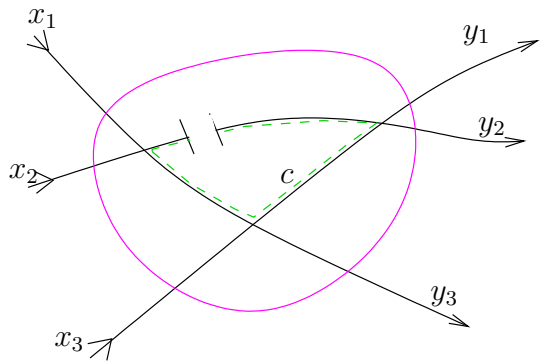}
    \caption{the scattering matrix associated to a cycle}
    \label{fig:scatt_cycle}
  \end{center}
\end{figure}

\end{demo}

 Taking the Logarithms,  we get 
the phase $ {\bf S}_h (\mu)/h =\sum _{j=0}^\infty S_j(\mu)
h^{j-1}  $ which is well defined modulo
a  multiple of $2\pi $.

\begin{lemm} Given the gauge transforms used in the normal form
of Theorem \ref{theo:normal},
the Taylor expansions of the $S_j$'s at $\mu=0$
are uniquely defined.
\end{lemm}
\begin{demo}
We will use the fact  that the local scattering matrix
computed in Section  \ref{sec:local}
is {\it irreducible} in the domain
$h^N \leq |\ga(\mu)| \leq \sqrt{h} $, meaning that none of the
 entries are
$O(h^\infty )$ in this domain.
It implies that, for each $j$,  the $W_{\pm, j}^{\rm in,~out}  $'s can be, 
up to a global multiplication by a symbol, defined as sets of WKB
solutions for which the scattering matrix is given by
Equations   (\ref{equ:T}) and (\ref{equ:AB})
with the value of  $\ga$ given by the normal form at
the point $z_j$.

\end{demo}

\section{The classical part $S_0$}\label{sec:S0}

\subsection{$S_0$ is classical}
We have the following:
\begin{theo}For any simple cycle $c$ of $C_0$, the function 
$S_0(\mu )$ depends only on the dispersion relation $C_\mu$.

\end{theo}
\begin{demo}
From the definition,  $S_0$ depends only on the
terms in $1/h$ in the phases of the images by our normal form transformations
of the explicit solutions of the normal form.
Those terms depends only on the  canonical
transformations used in the normal form and the associated
 generating
 functions
via stationnary phases (the Lagrangian manifolds).
\end{demo}

\subsection{$S_0$ as a regularized action integral}

Let us denote by $|\mu |=d(\mu , Z )$.
As in \cite{CP3}, it would be nice to get $S_0(\mu )$ as
a regularisation of an usual action integral. A basic fact in
\cite{CP3} was that any simple  cycle $c $ is a limit of a cycle
of $C_\mu$ as $\mu \ra 0_{\pm }$.
This is no longer the case here because
$\gg _0 \geq 0 $; one can  see an example in section \ref{ex}.
The idea is now to forget the initial problem and to work only with
the dispersion relation $C_\mu $ which can be embedded into
a larger family 
$C_{t,\mu}$ for which we can define action integral in some
suitable sectors of the $(t,\mu)$ space. We can then restrict
to $t=0$ and get our actions $S_0$.

We will  calculate $S_0$ by first computing
the same object $\Sigma _0(t, \mu) $  for
$C_{t,\mu}=\{ p(x,\xi, \mu)-t=0 \} $ and 
taking $S_0(\mu) =\Sigma _0(0, \mu) $.

The cycle $c$   is  a  limit of a cycle
$c(t,\mu) $ of $C_{t, \mu}$ as $(t,\mu) \ra 0 $
in some sector $\Omega _\pm :=\{ (t,\mu)| \pm t > 0,|\mu | << |t|\}$ .
 
We have, for $(t, \mu) \in \Omega$,
\[ \Sigma _0 (t,\mu)=
\int _{c_{t,\mu}}\xi dx +\sum _{j=1}^p  \pm \gg _{0,j} (t,\mu)
 (\ln |\gg_{0,j} (t,\mu)| -1) ~,\]
where the $\pm $ signs depends on orientation and 
can be determined from the Logarithmic singularities
of the action integrals.
The contributions $ \pm \gg _{0,j} (t,\mu)
 (\ln |\gg_{0,j} (t,\mu)| -1) $ come from the phase
shift between 
$x_+ ^{i\gg /h}$ and $\xi_+^{-i\gg/h}$ expressed as a
WKB function of the single variable $x$.

Knowing that $\Sigma _0$ is smooth, the previous formula
 defines the Taylor expansion
of $\Sigma _0$ w.r. to $(t,\mu)$
and hence the Taylor expansion of $S_0$ w.r. to $\mu$.

\subsection{The analytic case}

In the analytic case, we could also consider the Riemann surfaces
$X_\mu= \{ p_\mu =0\}$ and look at some complex cycles $c_\mu $ on
$X_\mu$ whose limit is $c$. Those cycles are not unique, but the 
real part of their  action
integrals are well defined and we can then take directly
the previous regularisation.

\subsection{An example}\label{ex}

Let us consider the adiabatic equation:
\[  \frac{h}{i}\frac{dX}{dt}=A_\mu (t) X \]
with 
\[ A_\mu (t)=\left( 
 \begin{array}{cc}
t^2 & \mu \\ \mu    & 2-t^2 \end{array} \right)\]
and the only cycle $c_0$ of $C_0$ passing by the singular
points $(\pm 1, \pm 1 )$.
It is clear that $c_0$ is not a limit of real cycle
$c_\mu $ of $C_\mu $, because the matrix $A_\mu (t)$ has real eigenvalues
for each $t$ and so $C_\mu $ is the union of 2 disjoint graphs
and has no real cycle.

\section{The subprincipal action} \label{sec:S1}

We know that the Landau-Zener coefficient given by Equation
(\ref{equ:tau}) is $0(h^\infty )$ if $|\mu |>> \sqrt{h}$.
It implies that in order to solve our problem up to
$O(\sqrt{h})$ terms it is enough to know
$S_0$ mod $O(|\mu |^3)$ and $S_1$ for $\mu \in Z$.
Let us  assume that we have local coordinates so that
$0\in Z$.
We will describe below the calculus of $S_1(0)$. 

\begin{lemm}\label{lemm:bkw}
Assuming $|\mu|=0$, the principal part
$\vec{a}(x){\rm exp}(iS(x)/h) $ of the WKB solutions
of Equation (\ref{equ:1}) associated
to arcs $]z_j,z_{j+1}[$ of $C_0$
can be smoothly extended beyond  the singular vertices
as WKB functions.
\end{lemm}
\begin{demo}
The property is invariant by FIO and it is enough to prove
 it for the
solutions of the 
normal form given in Equation (\ref{equ:sol_nf}).
The point  is that $a_0(0)=0 $,
hence $\gg =O(h^2)$. \end{demo}

The previous result is related to the fact  that the adiabatic
 theorem
 is still 
valid in case of eigenvalue crossings (see \cite{A-E,B-F}).

We will define on $c$ a piecewise smooth Hermitian line bundle $L$ with a
 connection
as follows:
\begin{itemize}
\item On each arc $[z_j, z_{j+1}]$,
$L_z= \ker H_{\rm class} (z) $ with the connection given by the transport
equation as in \cite{E-W} (in the
case of the adiabatic limit, it is the so called {\it geometric
 connection}
or {\it Berry phase}  \cite{Be}).
\item At each singular point, there are 2 limit fibers
$L_{\pm, j}$ and from $A_{0}$ (defined in Remark \ref{rem:unique})
we have an isomorphism between both limits given by transporting
the isomorphism
$(1,0)\ra (0,1)$
of $\C \oplus 0$ on $0\oplus \C $ by $(A_{0})_{|\C^2 \oplus 0}$.
\end{itemize}
\begin{defi} 
 The  phase ${\rm exp}(i S_1^\nabla  (0))$
is the holonomy of the  discontinuous line bundle $L$.
\end{defi}

\begin{figure}[hbtp]
  \begin{center}
    \leavevmode
    \input{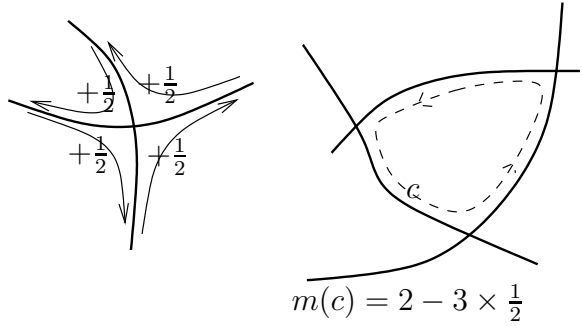}
    \caption{the singular Maslov indices}
    \label{fig:maslov}
  \end{center}
\end{figure}
Using the calculus of \cite{CVN} (page 20)
(we alert the reader that the previous convention 
for Maslov indices are not the same in the paper
\cite{CP3}), we can also  put the:
\begin{defi}
The (singular) Maslov index $m(c)\in \Z/2$  of a simple cycle $c$
which is the boundary of a bounded connected component
of $T^\star \R \setminus C_0$ is given by:
$m(c)=m_{\rm smooth}(c)+m_{\rm sing}(c)$
where $m_{\rm smooth}(c)$ is the usual Maslov index of a smooth
deformation
of $c$ while $m_{\rm sing}(c)$ is a sum of $\pm \ha$ 
associated to the singular points according to the rules
of Figure \ref{fig:maslov}.

The Maslov index of any cycle is defined by linearity from the
 previous Maslov
indices, so it gives a {\it cocycle.}
 For example, the Maslov index of a smooth cycle (even if not
simple) is the usual one, namely $\pm 2$.

\end{defi}

\begin{theo}
Using the previous definitions, we have:
\[ S_1(0) =S_1^\nabla  (0)+ m(c)\frac{\pi}{2}~. \]
\end{theo}
\begin{demo}
The proof follows essentially the lines of \cite{CP3} p. 474-476.

Let us give some details. A priori, there are several cases to check depending
on the position of the cycle $c$ at the singular points w.r. to the 
verticals. We will assume that the matrix
\[ \chi ' (O)= \left( \begin{array}{cc}
a&b \\ c& d \end{array} \right) \]
of the canonical transformation $\chi =\chi _0 $
satisfies $a\ne 0$ and $b \ne  0$, this is the {\it generic case.}
We define
\[ \ge = \left\{ \begin{array}{c}+1 {~\rm if~}ab >0\\
-1  {~\rm if~}ab <0 \end{array} \right.  ~.\]
The generating function
$\gf (x,y)= \gf _2 (x,y)+O(|x|^3 + |y|^3)$
of $\chi $, defined by $\chi (y,-\pa _y \gf)=(x,\pa _x \gf )$,
satisfies $\gf _2 (x,y)=\frac{1}{2b}(dx^2-2xy+ ay^2)$.
We need to compute mod $o_h(1)$
 the values for $x$ close to $0$  of the images 
by the normal form transform of 
\[ W_-^{\rm out}(y)=
\left( \begin{array}{c}Y(y)  \\ 0  \end{array} \right) \]

and 
\[ \widehat{W_-^{\rm in}}(\eta)=
\left( \begin{array}{c} 0 \\ Y (\eta)  \end{array} \right) ~. \]

Let us assume that the principal symbol of the $\Psi DO$ gauge transform
is the  $N\times N$ matrix
\[ \gs (A)(y,\eta)= \left( \begin{array}{ccc}
\vec{\ga}_1 (y,\eta) & \vec{\ga}_2 (y,\eta) &\cdots 
 \end{array} \right)~. \]

We get for the components of both WKB solutions for $x$ small but nonzero:

\[ W_{-,j}^{\rm out}(x)=
(2\pi h )^{-3/2}
\int _{y'\geq O}e^{\frac{i}{h}(\gf(x,y)+(y-y')\eta )}
C(x,y) \vec{\ga}_1(y,\eta)
 dy dy' d\eta ~,\]
with $C(0,0)=|b|^{-\ha}$,
and
\[ W_{-,j}^{\rm in}(x)=
(2\pi h )^{-1}
\int _{\eta \geq O}e^{\frac{i}{h}(\gf(x,y)+ y\eta )}
C(x,y)\vec{\ga}_2(y,\eta)
 dy  d\eta ~.\]

If we evaluate the integrals  by stationnary phase, the dominant contributions
come from the critical points and not from the boundary.
The determinant of both Hessians are the same, while the signature
differs by $1$.
The final result follows then by
\begin{itemize}
\item
 Looking at the value of the stationnary
phase calculations as $x$ is close to $0$:
the limits are respectively
 \[C(0,0)e^{i\ge \pi /4}  \vec{ \ga}_1(0,0)
 \]
and
 \[ C(0,0) \vec{\ga}_2(0,0)
~ \]
\item if $\ge >0$, one should add a contribution of the smoothed $c$,
while if $\ge <0$ there is no such contribution.
\item Remembering  that 
\[ {A_0}_{|\C^2 \oplus 0}= \left( \begin{array}{cc} \vec{\ga}_1(0,0)& \vec{\ga}_2 (0,0)
\end{array} \right)~. \]
\end{itemize}

\end{demo}

\section{Application 1: adiabatic limit with avoided crossings}
\subsection{Adiabatic limit}
\label{sec:adiab}

We consider the following equation:
\begin{equation} \label{equ:Adia}
 \frac{1}{i}\frac{dX}{d\tau}=A_\mu (h\tau) X \end{equation}
where  $A_\mu (t),~0\leq t \leq a, $
is a self-adjoint matrix which is  smooth w.r. to 
$(t, \mu)$ and we consider  $0\leq \tau \leq a/h $.
The limit $h \ra 0 $ of this equation is called the {\it adiabatic
limit.}

We can rewrite Equation (\ref{equ:Adia}) in a standard {\it semi-classical}
form  by puting
$t=h\tau$:
\begin{equation} \label{equ:adia}
 \frac{h}{i}\frac{dX}{dt}=A_\mu (t) X \end{equation}
where $0\leq t \leq a$.

We will assume that the eigenvalues of $A_\mu(0) $
and $A_\mu (a) $ are all non degenerate.
The scattering matrix ${\cal S}(\mu, h): \C ^N \ra \C ^N $
is defined by $X(0) \ra X(a) $ where $X$ is a solution of 
Equation (\ref{equ:adia} ).

\subsection{Outside eigenvalues crossings}

Let $\gl (t)$ be an eigenvalue of multiplicity $1$ of 
$A_0(t)$ for $t$ in some open intervall $I$.
Then Equation (\ref{equ:adia}) admits a unique (up to 
multiplication by some function of $h$)
formal {\it WKB solution} given by
\[ X(t)=e^{i\Lambda (t)/h}\left(\sum _{j=0}^\infty a_j (t) h^j \right) \]
where
$\Lambda ' (t)= \gl (t) $ and
 $a_0(t)$
satifies:
\begin{itemize}
\item $a_0(t) \in \ker \left(A_0(t)-\gl (t)\right) $
\item $\nabla a_0 (t) =0 $
where $\nabla $ is the {\it geometric} or {\it Berry } connection.
\end{itemize}

Let us recall that $\nabla _{\pa / \pa t}a (t)= \Pi_t  a'(t) $
where $\Pi _t $ is the orthogonal projection of $\C ^N $
onto the eigenspace $\ker (A_0(t)- \gl(t) )$.

The previous statement is the content of the so called {\it quantum
adiabatic theorem} and goes back to \cite{B-F}.

\subsection{Avoided crossings}

What happens when eigenvalues become degenerate at some values
of $t$?

Let us try to understand the {\it generic situation}.
It is well known that eigenvalue crossings for a real symmetric
(resp. complex Hermitian) matrix is a codimension $2$ (resp. $3$)
property. It is the content of the well known 
Wigner-Von Neumann theorem \cite{WvN}.

Physically, eigenvalues crossings can still occur for symmetry
reasons. But, if we break the symmetry by a small perturbation of 
size $\mu$,
we will get the  so-called  {\it avoided crossings.}
We have now two small parameters: the semi-classical (adiabatic)
parameter $h$ and the perturbation parameter $\mu$.
The previous results allow to discuss the uniform expansion
of the scattering matrix w.r. to both small parameters.
\begin{figure}[hbtp]
  \begin{center}
    \leavevmode
    \input{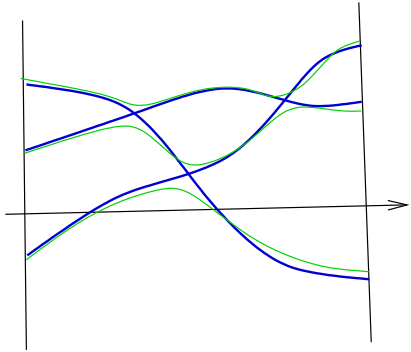}
    \caption{the dispersion relation for the adiabatic limit}
    \label{fig:adia}
  \end{center}
\end{figure}

\subsection{Precise assumptions}

We will assume that the eigenvalues of $A_0(t)$ cross
 transversally 
only by pairs on $]0,a[$.
The dispersion relation $C_\mu \subset T^\star [O,a  ] $
is defined by
 $p_\mu (t, \tau) ={\rm det}(\tau {\rm Id} - A_\mu(t))$.
So that $C_\mu$ is exactly the union of the graphs
 of the eigenvalues
of $A_\mu(t)$.

\subsection{Calculation of the scattering matrix}

\begin{figure}[hbtp]
  \begin{center}
    \leavevmode
    \input{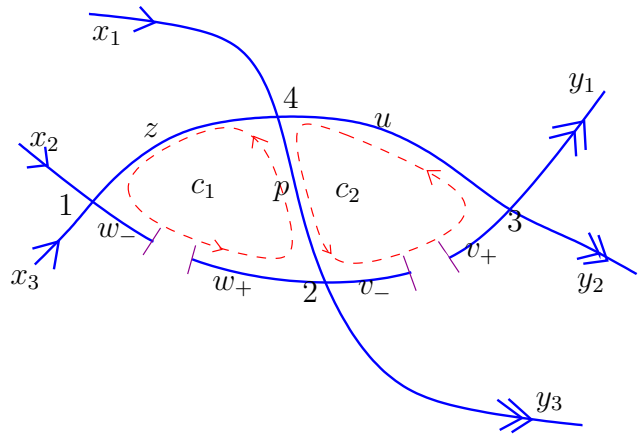}
    \caption{recipe for the global scattering matrix}
    \label{fig:global}
  \end{center}
\end{figure}

Let us describe how to compute the global scattering matrix
in the case of Figure \ref{fig:global}.
Let us start with the 4 local scattering matrices
${\cal S}_j,~j=1,\cdots, 4$ and the 2 holonomies
${\cal H}_\mu (c_k),~k=1,2$.

We try to describe a global solution of our system
which is given from WKB solutions associated to each arc
of a maximal tree of $C_0$.
We have 10 equations with 13 unknowns which allow to compute
$\vec{y}$ from $\vec{x}$.

\[ \left\{ \begin{array}{l}
w_-={\cal H}_\mu (c_1) w_+  \\
v_-={\cal H}_\mu (c_2) v_+  \\
 \left( \begin{array}{c} z \\ w_- \end{array} \right)
= {\cal S}_1 
 \left( \begin{array}{c} x_3 \\ x_2 \end{array} \right) \\
 \left( \begin{array}{c} v_- \\ y_3 \end{array} \right)
= {\cal S}_2 
 \left( \begin{array}{c} w_+ \\ p \end{array} \right) \\
 \left( \begin{array}{c} y_1 \\y_2  \end{array} \right)
= {\cal S}_3 
 \left( \begin{array}{c} v_+ \\ u \end{array} \right) \\
 \left( \begin{array}{c}u \\ p \end{array} \right)
= {\cal S}_4 
 \left( \begin{array}{c} z \\ x_1 \end{array} \right) \\
\end{array}
\right. \]
It turns out that the global scattering matrix is the product 
of 5 unitary matrices as follows~:
\[ \vec{x} \ra \left( \begin{array}{c}x_1 \\ z \\w_-
\end{array} \right) 
\ra \left( \begin{array}{c}u \\ p \\ w_+
\end{array} \right) 
\ra \left( \begin{array}{c}u \\v_- \\ y_3
\end{array} \right) 
\ra \left( \begin{array}{c}u \\v_+ \\ y_3
\end{array} \right)
\ra \vec{y}~. \]

\section{Application 2: EBK quantization rules}

\begin{figure}[hbtp]
  \begin{center}
    \leavevmode
    \input{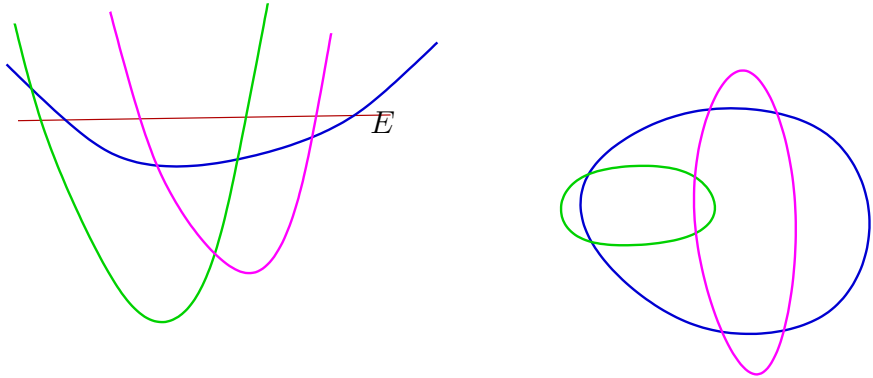}
    \caption{the dispersion relation for the Born-Oppenheimer
Hamiltonian}
    \label{fig:ebk}
  \end{center}
\end{figure}
We consider a Born-Oppenheimer
Hamiltonian of the following form:
\[ \widehat{K_\nu}=
-h^2 \frac{d^2}{dx ^2} \otimes {\rm Id}
+V_\nu (x) \]
where 
$V_\nu : \R \ra {\rm Herm}(\C ^N )$
is smooth w.r. to $(x,\nu)$.
We assume:
\begin{itemize}
\item
 The eigenvalues of $V_0(x)$ are of multiplicities
at most $2$ and cross transversally.
\item 
The following properness condition:
\[ V_\nu (x) \geq p(x){\rm Id} \]
where $\lim _{|x|\ra \infty } p(x)=+\infty $.
\item
We choose $E$ so that, for any $x \in \R$,
 $E$ is not a degenerate eigenvalue
of $V_0(x)$.
\item If the eigenvalue $\gl_j(x)$
of $V_0(x)$ satisfies  $\gl_j(x_0)=E$, then
$\gl_j' (x_0) \ne0$.
\end{itemize}

We can apply the previous method in order to compute EBK
 quantization
rules for the equation  $ (\widehat{K_\nu}-E) \vec{U}=O(h^\infty )
~.$

EBK quantization can be solved following the same path; but is
 this
case we have the
same number of equations than of unknowns and EBK rule
is given by the vanishing of a suitable determinant as in
 \cite{CP3}.

\bibliographystyle{plain}

\end{document}